\documentclass{article}
\usepackage{verbatim}
\usepackage{float}
\usepackage{mathtools}
\usepackage{enumerate}
\usepackage{xargs}[2008/03/08]
\usepackage{mathrsfs}

\usepackage[cp1251]{inputenc}
\usepackage[english]{babel}
\usepackage{amsfonts,amssymb,amsmath,amsthm,euscript,xspace,cite}
\usepackage{bbm}   
\usepackage{a4, graphicx}
\title{On Universality of Regular Realizability Problems\thanks{This work is supported by the Russian Science Foundation grant 20--11--20203.}}

\author{Michael Vyalyi\thanks{Faculty of Computer Science, National Research University Higher School of Economics,
 Pokrovsky boulevard 11,   Moscow, 109028,  Russia,
 \texttt{vyalyi@gmail.com}
  } \and Alexander Rubtsov\thanks{Faculty of Computer Science, National Research University Higher School of Economics,
 Pokrovsky boulevard 11,   Moscow, 109028,  Russia,
 \texttt{rubtsov99@gmail.com}}
}

\date{}

\def\NN{\mathbb N}
\def\FF{\mathbb F}
\let\epsilon\varepsilon
\let\eps\varepsilon
\let\al\alpha

\let\ph\varphi

\let\ld\lambda

\def\sm{\setminus}
\def\es{\varnothing}

\theoremstyle{plain}
\newtheorem{theorem}{Theorem}

\newtheorem{lemma}[theorem]{Lemma}
\newtheorem{prop}[theorem]{Proposition}
\newtheorem{cor}[theorem]{Corollary}
\theoremstyle{definition}

\newtheorem{ex}{Example}
\let\la\langle
\let\ra\rangle
\let\epsilon\varepsilon
\let\eps\varepsilon
\let\al\alpha

\let\ph\varphi

\let\ld\lambda

\def\bb{{\vartriangleleft}}
\def\eb{{\vartriangleright}}
\def\bv{\boldsymbol v}
\def\bi{\boldsymbol i}

\def\F{\ensuremath{\mathcal F}}
\def\Fg{\ensuremath{{\mathcal F}^g}}
\def\C{\ensuremath{\mathscr C}}

\def\A{\ensuremath{\mathcal A}}
\def\I{\ensuremath{\mathcal I}}

\def\H{\ensuremath{\mathcal H}}

\def\B{\ensuremath{\mathcal B}}

\def\Y{\ensuremath{\mathcal Y}}

\def\U{\ensuremath{\mathcal U}}
\def\R{\ensuremath{\mathcal R}}
\def\G{\ensuremath{\mathcal G}}
\def\Sc{\ensuremath{\mathcal S}}

\def\tu{\tilde u}

\def\bv{\boldsymbol v}

\def\poly{\mathop{\mathrm{poly}}\nolimits}  

\def\leGen#1#2{\mathop{\leq^{\mathrm{#2}}_{\mathrm{#1}}}}

\def\lelog{\leGen{m}{log}}
\def\leP{\leGen{m}{P}}

\def\lerat{\mathop{\leq_{\mathrm{rat}}}}

\def\lettP{\leGen{dtt}{P}}
\def\lettNP{\leGen{dtt}{FNP}}

\def\leTNP{\leGen{T}{FNP}}

\def\reg{\mathrm{DRR}}
\def\nreg{\mathrm{NRR}}

\let\es\varnothing

\newcommand*\FNP{\ensuremath{\mathrm {FNP}}}
\newcommand*\NL{\ensuremath{\mathrm {NL}}}
\newcommand*\FNL{\ensuremath{\mathrm {FNL}}}

\renewcommand*\P{\ensuremath{\mathrm {P}}}
\newcommand*\NP{\ensuremath{\mathrm {NP}}}

\newcommand*{\size}{\mathop{\mathrm{size}}}

\newcommand*\PSPACE{\ensuremath{\mathrm {PSPACE}}}

\usepackage[normalem]{ulem}
\usepackage{color}
\usepackage{xcolor}

\usepackage{enumerate}

\usepackage{marginnote}

\definecolor{WildStrawberry}{HTML}{FF43A4}
\definecolor{DarkGray}{HTML}{666666}

\usepackage{ifthen}

\newcommand{\malert}[2][]{\marginpar{\color{WildStrawberry}\ifthenelse{\equal{#1}{}}{\vspace{-2ex}}{\vspace{-2ex}\vspace{#1}}  #2}}

\newcommand{\mcomment}[2][]{\marginpar{\ifthenelse{\equal{#1}{}}{\vspace{-2ex}}{\vspace{-2ex}\vspace{#1}} \color{DarkGray} #2}}
\newcommand\reduline{\bgroup\markoverwith
  	{\textcolor{red}{\rule[-0.5ex]{2pt}{0.4pt}}}\ULon}

\usepackage{dashrule}

\newcommand\alertuline{\bgroup\markoverwith
  	{\color{WildStrawberry}{\hdashrule[-0.5ex]{2pt}{0.4pt}{1pt}}}\ULon}

\newcommand\commentuline{\bgroup\markoverwith
   		   	{\color{DarkGray}{\hdashrule[-0.5ex]{2pt}{0.4pt}{1pt}}}\ULon}

\begin{document}
\maketitle

\begin{abstract}
  We prove the universality of the regular realizability problems for
  several classes of filters. The filters are encodings of finite
  relations on the set of non-negative integers in the format proposed
  by P. Wolf and H. Fernau. The universality has proven up to
  disjunctive truth table polynomial reductions for unary relations and polynomial space reductions for invariant binary relations. Stronger reductions correspond to the results of P. Wolf and H. Fernau about decidability of regular realizability problems for many graph-theoretic properties.
\end{abstract}

\section{Introduction}

The class of regular languages is an important class in Formal Language Theory. It is well-known that basic algorithmic problems for regular languages are decidable (in particular, the emptiness problem, the finiteness problem, the membership problem). Moreover, they are decidable in polynomial time if an instance of a problem is an encoding of the transition relation of a finite automaton recognizing the language.

Checking regular conditions on words of an arbitrary language can be harder. We refer to the corresponding problem as the \emph{regular realizability} problem. Regular realizability problems form a class of algorithmic problems parametrized by languages. 	 Fix a~language $F$ called a \emph{filter}. An instance of the \emph{regular realizability problem} $\reg(F)$ ($\nreg(F)$) is  a regular language $L(\A)$ described by a deterministic finite automaton (DFA)  (by a nondeterministic finite automaton (NFA))~$\A$  respectively. The problem is to verify the non-emptiness of the intersection  $F\cap L(\A)$.

Regular realizability problems can be viewed as reachability problems on graphs with restrictions on paths. More definitely, assume that edges of a graph are labeled by symbols and restrict to the paths such that a word read along a path belongs to a specified language (the filter). For example, the CFL restrictions were intensively studied (see~\cite{Ya90, BEM97, RV15, CMS22}). Note that in this case the regular realizability problem is decided in polynomial time.

General form of regular realizability problem was suggested in~\cite{Vya11}.  That paper provides  examples of regular realizability problems that are complete for many complexity classes and for the class of recursively enumerable languages.

Generalizing the question of the existence of algorithmically hard problems, we come to a question about `density' of a certain class of problems  among all possible decision problems. For this purpose we identify an algorithmic problem with a language of instance descriptions giving a positive answer to the problem. This results in a class of languages. The question is about an approximation of an arbitrary language by  languages from this restricted class. Since we are interested in algorithmic complexity, it is natural to use (quasi-)orders of algorithmic reductions between languages instead of metrics or measures on language classes.

Reductions are the main tool in computational complexity theory. Any reduction is a binary relation $A\leq B$ on languages, which is a quasi-order. To make  a question about `density' of a language class formal, we introduce a \emph{universality property} of a language class $\C$ relative to reductions $(\leq_1, \leq_2)$ as follows: for any non-empty language $X$ there exists a language $L_X\in \C$ such that
\[
X\leq_1 L_X\leq_2 X.
\]
In this way, we get a vast family of universality properties. Weaker reductions correspond to stronger properties. From an algorithmic point of view, it is reasonable to restrict  attention to reductions that are not stronger than Turing reductions. For applications to computational complexity, weaker reductions are needed (see examples below, Section~\ref{ssec:reductions}).

In~\cite{Vya13},  the universality property of DRR relative to reductions  $(\lelog,\leGen{dtt}{\mathrm{FNL}}) $  was proved.
Here the first reduction is the $m$-reduction in logarithmic space (see below Section~\ref{ssec:reductions}), the second one is the disjunctive truth table reduction by functions computable with an  $\NL$-oracle (for disjunctive truth table reductions see Section~\ref{ssec:reductions}; for the class \FNL{} see~\cite{Vya13}, here we just note that all functions from this class are computable in polynomial time). It was also proved in~\cite{Vya13} that in each class of polynomial hierarchy there exists a complete DRR problem.  Note that the results of~\cite{Vya13} can be easily reproduced for NRR.

The universality property is meaningful for the regular realizability problems with restricted filters. Results of this type can be interpreted as a universality of a certain computational model. In~\cite{Vya13} the universality property of DRR relative the same reductions was proved for prefix-closed filters. It means a universality of automata using a general nondeterminism (see~\cite{Vya09}). This results was extended in \cite{RV22}: the universality of NRR relative to reductions $(\lelog, \leGen{T}{P})$ was proved for the filters that are correct protocols of work of finite automata equipped with an auxiliary data structure. Here  $\leGen{T}{P}$ is a polynomial time Turing reduction. This result implies that any sufficiently strong complexity class can be described as a~class of languages recognizable by Turing machines operating in logarithmic space and equipped  with a suitable auxiliary data structure.

In this paper, we clarify the boundary between filter classes with and without universality properties. And now the motivation for choosing a filter class is different and is not related to computation models.

Regular realizability problems (under the name regular intersection emptiness problems) was studied in~\cite{WolfFernau20, W22} for other reasons. The basic question is the following. Fix a graph-theoretic property and a format of graph encodings by words over a finite alphabet. The question is to check that there exists a graph such that its encoding satisfies some regular condition. In other words, an instance of the problem is a set of graphs. The graphs in the set have encodings satisfying a~regular condition, i.e. an input of  $\nreg(F)$. We are interested in the existence of a~graph in the set satisfying the property described by a filter $F$. In these settings, a~crucial step is a~choice of the encoding format. It is a common issue in combining formal languages and computational complexity theory: how to encode more complicated structures (graphs, trees, etc.) by words over a finite alphabet. Usually, there is no natural choice of an encoding.  A~very interesting format for graph encodings was suggested in~\cite{WolfFernau20}. In this format vertices are integers represented in unary and edges are pairs of integers. The most important features of this format are (i) an  order in a list of edges is arbitrary and (ii) repetitions of edges in the list are allowed. It makes regular realizability problems easier since a fixed order in the list causes additional difficulties (see, e.g.~\cite{RV22}, where the order of protocol blocks is essential).

Using this format, for many typical graph-theoretic properties, the decidability of regular realizability problems was proved in~\cite{WolfFernau20}. This line of research was continued in~\cite{W22, DFW21}.

Informally speaking, the results in~\cite{WolfFernau20, W22, DFW21} show that this specific format of graph encodings is structurally easy. It implies that the universality for filters in the form of graph encodings is harder to achieve.

Nevertheless, in this paper we present universality results for graph encodings and for encodings of finite unary relations  on the set of nonnegative integers. These results use stronger reductions than the general universality theorem. This reflects the specific properties of this class of RR problems.

The main obstacle in achieving universality results for these filters is freedom in listing elements of a set. To overcome this difficulty we apply efficient asymptotically good codes---an exotic construction for Formal Language Theory. Also, we introduce a new model of Boolean functions computation. Boolean functions and families of subsets of a fixed finite set are essentially the same, and the new model is defined using families of subsets. The complexity measure for this new model generalizes the size of decision trees and branching programs. Here we do not study the model systematically. Note that, from this point of  view, we get  examples of Boolean functions such that their complexity in this model is polynomially lowerbounded by the number of points in which a~function takes the value~1.
Unlike the above examples of computation models, in this model the computation of a~function value at a point is  $\NP$-complete (see corollary~\ref{Gell-value} below), but it is possible to construct a list of all points in which a~function takes the value~1 in time polynomially bounded by the number of such points (see lemma~\ref{lm:next-label} below).

The rest of the paper is organized as follows. In Section~\ref{sec:def} we give exact definitions for the questions outlined above and recall basic definitions for finite automata and algorithmic reductions. In particular, we provide  the exact definition of the encoding format of finite relations on $\NN$ of fixed arity.\footnote{In this paper we mean that $\NN$ is the set of non-negative integers.} Up to  minor modifications, it is a straightforward generalization of the definition from~\cite{WolfFernau20}. The modifications do not affect the algorithmic complexity of problems considered here. In Section~\ref{sec:automata-infinite-alphabet} we introduce a technical tool to analyze RR problems for relation encodings. We change the usual automata by the automata over the infinite alphabet $\NN^k$, here $k$ is the arity of relations. For finite encodings of automata of this kind we rely on structural results about regular languages over the unary alphabet, namely, on the Chrobak-Martinez theorem.

In Section~\ref{sec:sketch} we present a general approach for achieving the universality results. To realize this approach we introduce in Section~\ref{sec:sets-by-graphs} the problem of representing finite subsets of a fixed set (`the universum') by directed graphs. We prove in this section basic results on the sizes of graphs representing  families of subsets with the separability property (in other words,  families with the low intersection property).  In the construction of families with  the separability property we need efficient asymptotically good codes.

Since the main motivation is to study graph-theoretic properties, we separately analyze relations that are invariant under bijections of $\NN$. Indeed, a typical graph-theoretic property is invariant under isomorphisms. Universality results for invariant relations are harder to achieve. We prove the universality relative to disjunctive truth table reductions in polynomial space.

The next three sections contain the exact statements and proofs for universality results. In Section~\ref{sec:ubu} we prove universality for encodings of finite unary relations (i.e.\ finite subsets of $\NN$). In Section~\ref{sec:bbu}  we prove universality for  finite binary invariant relation encodings. In Section~\ref{sec:uuu} we prove the universality for  finite unary invariant relation encodings. This last result is proved for languages in the unary alphabet. This restriction on the alphabet seems inevitable, since the only invariant of a finite unary relation under bijections of $\NN$ is its size and  it is  impossible  to encode binary words by reasonably short unary words.

In the final section we discuss the results obtained and  possible directions of future research in this area.

  \section{Definitions and Basic Constructions}\label{sec:def}

  We recall in this section the basic definitions for automata, transducers, complexity classes and algorithmic reductions and fix notation to be used in the sequel. Also we give in this section definitions related to the  format of  finite relation encodings.

  \subsection{Automata and  transducers}

A (nondeterministic)  automaton $\A$ over an alphabet  $\Sigma$ is specified by the following data: a finite state set $Q$, the initial state $q_0$, the set of accepting states $Q_a$, the transition relation $\delta_\A\subseteq Q\times (\{\eps\}\cup \Sigma) \times Q$, where $\eps$ is the empty word. The transition relation is extended to the relation  $\delta_{\A}\subseteq Q\times \Sigma^*\times Q$ by the rule: $(q_1, w a, q_2)\in \delta_\A$ is equivalent to the existence of $q_3$ such that  $(q_1, w , q_3)\in \delta_\A$ and $(q_3,  a, q_2)\in \delta_\A$. Here $w\in\Sigma^*$, $a\in \Sigma\cup\{\eps\}$. The elements of the transition relation will be also written as
$
q_1\xrightarrow{w} q_2\,.
$
If needed, we will join arrows as it shown below:
\[
q_1\xrightarrow{w_1} q_2\xrightarrow{w_2} q_3
\quad\text{is equivalent to}\quad
(q_1\xrightarrow{w_1} q_2)\land(q_2\xrightarrow{w_2} q_3)\,,
\]
which implies  $q_1\xrightarrow{w_1w_2} q_3$\,.

We allow infinite alphabets in the definition of automata. This non-standard definition is convenient for our purposes. If the alphabet is finite, then an automaton can be specified by a list of transitions. For infinite alphabets, an encoding of the transition relation should be specified  (see below an example in Section~\ref{sec:automata-infinite-alphabet}).

An automaton  is called  \emph{deterministic}  if, for each pair  $(q,a)$, $q\in Q(\A)$, $a\in \Sigma$, there exists at most one $q'$ such that $(q, a, q')\in \delta_\A$, and, if there exists a~transition $(q,\eps, q')$, then there is no transition $(q,a, q'')$, $a\in \Sigma$.

A word $w$ is \emph{accepted} by an automaton $\A$ if $(q_0, w , q_a)\in \delta_\A$ for $q_a\in Q_a$. A~\emph{language} is a subset of words over a finite alphabet. The language $L(\A)$ is \emph{recognized} by an automaton $\A$ if it consists of all the words $w$ accepted by $\A$.
Automata  $\A'$ and $\A''$ are \emph{equivalent\/} if $L(\A')= L(\A'')$.

Transitions in the form  $(q_1,\eps, q_2)$ are called $\eps$-transitions. Given an automaton $\A$, it is possible to construct an equivalent automaton $\A'$ without $\eps$-transitions. For each non-$\eps$-transition $(q_1, a, q_2)\in\delta_{\A}$ the automaton $\A'$ has transitions $(q_3, a, q_4)$, where $q_1$ is reachable from $q_3$ by $\eps$-transitions and $q_4$ is reachable from $q_2$ by $\eps$-transitions. This transformation takes a polynomial time in the size of $\A$ since it is reduced to the reachability problem  for finite directed graphs. So  we assume by default that automata have no  $\eps$-transitions and apply this transformation, if necessary.

A \emph{run} of an automaton $\A$ on  a word $w= w_1w_2\dots w_n$ is a sequence of states $q_0, q_1, \dots q_n$ such that $(q_{i-1}, w_i, q_i)\in \delta_A$ for all $1\leq i\leq n$.  A run from the initial state to an accepting state is called  \emph{accepting}.

Hereinafter we assume that automata in consideration do not have unreachable and dead states. It means that each state belongs to an accepting run of the automaton. Deleting unreachable and dead states is possible in polynomial time since it is also reduced to the reachability problem for finite directed graphs. Such automata are known as \emph{trimmed}.

     A \emph{finite state transducer} (FST) is a nondeterministic
     finite automaton with an output tape. In addition to the transition relation, for each transition a finite set of words over the \emph{output alphabet}  is specified. Making a transition, the transducer should  append a word from this set to the content of the output tape. An FST  $T$ defines a binary relation on words: $uTv$ holds if there exists an accepting run of $T$ on the input $u$ such that at the end of the run the word~$v$ is written on the output tape. The \emph{rational dominance} relation $A \lerat B$ on languages holds  if there exists an FST~$T$ such that
\[A = T(B)= \{v \mid \exists u: uTv,  u\in B\}.\]

\subsection{Reductions}\label{ssec:reductions}

A language $A$ is  $m$-reducible to a language  $B$ by  functions from the class  $\C$ (notation $\leGen{m}{\C}$), if there exists a function $f\in\C$ such that  $x\in A$ is equivalent to  $f(x)\in B$. A reducibility relation is a preorder (i.e.\ a transitive and reflexive relation) if  $\C$ contains the identity function and is closed under compositions. We denote by   $\leP$ polynomial time reductions (Karp reductions) and by  $\lelog$ the reductions in log space.

Turing reductions (oracle reductions) are defined as follows:  $A\leGen{T}{\C} B$ if $A$ can be decided by  an algorithm that computes functions from the class   $\C$  relative to the oracle  $B$. (An algorithm decides $A$ if it computes the indicator function of $A$). Here we need the reduction  $\leTNP$, where functions from the class \FNP{} are computable in polynomial time w.r.t.\  the input size using an  $\NP$-oracle. An algorithm computing such a function can get in one step an answer to a question: whether a specified query word belongs to a language from  $\NP$. It is easy to see that an equivalent requirement is to check the membership to a fixed  $\NP$-complete language.

An  $\NP$-oracle gives an answer to the query. Thus an algorithm with $\NP$-oracle can decide the membership problem for an arbitrary $\NP$ language in one oracle call. Note that it is formally stronger than the usual definition of  $\NP$. In particular,  $\NP\ne\P^{\NP}$ if the polynomial hierarchy is infinite (it is  a standard conjecture in computational complexity theory).

We will need the following construction. Let  $R$ be a binary relation such that  $R \in \P$ and, for a polynomial $q(\cdot)$, it holds that $|y|\leq q(|x|)$ for all $(x,y)\in R$. Relations of this form are used in the definition of the class  $\NP$ and $y$ is called a  \emph{certificate} for $x$ if $(x,y)\in R$.
The existence of a certificate for  $x$ is exactly an $\NP$ condition. Using  $\NP$-oracle it is also possible in polynomial time to find a certificate for  $x$ provided that at least one certificate does exist.

For this purpose, it is sufficient to note that a language
 \[
 R_w = \{x : \exists z\; R(x,wz)\}
 \]
 (for a fixed word $w$) is in   $\NP $.  So an algorithm computes bits of a certificate one by one, using the $\NP$-oracle calls in the form
 $x\stackrel{\text{\tiny ?}}{\in} R_{w0}$, $x\stackrel{\text{\tiny ?}}{\in} R_{w1}$ provided  $x \in R_w $.
In the sequel, this algorithm will be referred to as the \emph{certificate construction algorithm}.

The truth table reductions are a restricted form of Turing reductions. In this case, the algorithm computes a list of queries $q_1,\dots, q_s$ to the oracle and applies a fixed Boolean function with  $s$ arguments to the oracle answers. The function determines the result of the computation. We will use even more restrictive \emph{disjunctive truth table reductions} (dtt reductions). In this case the result is the disjunction of oracle answers.

We need two types of dtt reductions. The stronger reduction  $\leGen{dtt}{\PSPACE}$ uses functions computable in polynomial space. Note that in this case the list of queries can be exponentially large w.r.t.\  the input size. This reduction makes sense for sufficiently hard languages. E.g., the corresponding preorder separates classes of the arithmetical hierarchy.

The weaker reduction  $\lettP$  uses functions computable in polynomial time to generate queries.

 \subsection{Finite Relation Encodings}\label{ssec:rel-descriptions}
 
 We are interested in filters that define finite relations (e.g.\ graphs). In this paper we consider only unary and binary relations. Nevertheless, the definitions will be done for the general case of $k$-ary relations.  Generalizing encodings from~\cite{RV22} and~\cite{WolfFernau20}, we use the following format. Let $R\subseteq \NN^k$ be a~finite $k$-ary relation.
A sequence $x = (x_1, x_2, \dots, x_k)\in \NN^k$ is encoded by a word $\la x\ra = \bb a^{x_1}\#a^{x_2}\#\dots\# a^{x_k}\eb$ over the alphabet $\{a, \bb, \eb, \#\}$. These words are called \emph{blocks}. An \emph{encoding} $\la R\ra$ of a~finite relation $R$ is any word in the form
\[
\prod_{i=1}^t\la x(i)\ra ,\quad\text{where}\  \{x(i): 1\leq i\leq t\} = R.
\]
Here  $\prod$ is the concatenation of the words. Note that the encodings of  elements of $R$ can appear in arbitrary order and repetitions are allowed. Thus there exist many encodings of the same relation.

Let us associate with the family of finite relations $\R$ the language
\[
\la\R\ra = \{\la R\ra : R\in\R\},
\]
that consists of all encodings of all relations from $\R$.
Languages $\la\R\ra$ are  filters for RR problems considered here.

\begin{ex}[Unary relations]
A unary relation $S\subseteq \NN^1$ is just a subset of $\NN$. The unary encoding of $S$ consists of words over the alphabet $\{a, \bb, \eb, \#\}$ in the form
\[
\bb a^{x_1}\eb\bb a^{x_2}\eb\dots \bb a^{x_t}\eb,
\]
where $\{x_i: 1\leq i \leq t\} = S$. For example, words
$\bb a\eb \bb aa\eb$ and $ \bb aa\eb \bb a\eb \bb a\eb$ are encodings of the set $ \{1,2\}$.
\end{ex}

Taking in mind the results of~\cite{WolfFernau20}, we are interested in encodings of simple undirected graphs without isolated vertices. We follow the rule suggested in~\cite{WolfFernau20} and assume that edges are listed in arbitrary order, the ends of an edge are listed in arbitrary order, blocks  $\bb a^x\# a^x\eb$ are allowed but just ignored. More exactly, a~binary relation $R$  defines a graph $G_R =(V,E)$ such that
\begin{equation} \label{eq:def:GR}
  \begin{aligned}
	&V = \{v : \exists u\; (u,v) \in R\} \cup \{u : \exists v \; (u,v) \in R\},\\
      	&E =\big\{\{u,v\}: (u\ne v)\land \big( ((u,v)\in R) \lor ((v,u)\in R)\big)\big\}.
  \end{aligned}
 \end{equation}
By definition, an encoding $\la G\ra$ of a graph  $G$ is an encoding of a relation  $R$ such that $G = G_R$.

\begin{ex}
  For the graph $G$ with vertices  $\{1,2,3\}$ and edges  $\big\{\{1,2\}, \{2, 3\}\big\}$ (a~path $P_3$), possible encodings are the words  \[ \bb a \# aa\eb\bb aa \# aaa\eb, \quad\bb aa \# a\eb\bb aaa \# aa\eb  \bb a \# aa\eb\bb aa \# aa\eb
  \] (the list is not complete, actually, there are infinitely many encodings since repetitions are allowed).
\end{ex}

Our encodings slightly differ from the encodings used in~\cite{WolfFernau20,DFW21, W22}.  We  modify the format of delimiters only. Thus one encoding can be transformed to another by a~FST. We are interested in algorithmic complexity of RR problems. As it is shown in~\cite{RV15},  $A\lerat B$ implies  $\nreg(A)\lelog \nreg(B)$. Therefore, these variances in encodings are insignificant.

Graph properties are usually assumed to be invariant under isomorphisms. We call a family of finite relations \emph{invariant} if it is closed under bijections of $\NN$. It follows directly from~\eqref{eq:def:GR} that encodings of graphs from a family closed under isomorphisms form an invariant family of relations.

\begin{ex}
  If  we are interested in an invariant family of graph encodings, then the path $P_3$ on 3 vertices has the following encodings
$\bb a \# aa\eb\bb aa \# aaa\eb$, \linebreak
$\bb aa \# aaa\eb\bb aaa \# aaaa\eb  \bb aaa \# aa\eb$,
$\bb aaaaa \# aa\eb\bb aa \# aaa\eb $
  (of course, the list is not complete).
\end{ex}

\begin{ex}
  Let $\H$ be a family of Hamiltonian simple graphs and  $\la\H\ra$ be the set of encodings of graphs from $\H$. Then
\[
\begin{aligned}
  &   \bb a \# aa\eb\bb aa \# aaa\eb\bb aaa \# a\eb\in \la\H\ra
  &&\text{(an encoding of the 3-vertex cycle),}\\
  &   \bb a \# aa\eb\bb aa \# aaa\eb\bb aa \# aaaa\eb\notin \la\H\ra
  &&\text{(an encoding of the star with   3 edges).}\\
\end{aligned}
\]
\end{ex}

\section{Automata over Infinite Alphabets}\label{sec:automata-infinite-alphabet}

Let  $\A$ be an automaton over the alphabet $\{a, \bb, \eb, \#\}$.
We denote by  $\R^k(\A)$ the class of  $k$-ary relations on $\NN$ whose encodings are accepted by $\A$, at least one encoding for each relation. Unary words are in one-to-one correspondence with  nonnegative integers, namely, a~word is mapped to the  length of the word. It is convenient in the further analysis to get rid of delimiters and consider automata over the infinite alphabet $\NN^k$.

A letter $\bi =  (i_1, i_2, \dots, i_k)$ of the infinite alphabet~$\NN^k$ corresponds to the word $\bb a^{i_1} \# a^{i_2}\# \dots \# a^{i_k}\eb$. The \emph{content} of a word $\bi_1\ldots \bi_n$ (i.e.\ the set of $k$-tuples $\bi_r$, $1\leq r\leq n$) is the relation encoded by the set $\{\bi_1, \ldots, \bi_n\}$.
For an automaton $\A$, we construct the automaton $\tilde\A_{\NN}^k$ having the same state set, the same initial state, the same accepting states as $\A$. The alphabet of $\tilde\A_{\NN}^k$ is $\NN^k$. The transition relation is defined as follows: for all $\bi =  (i_1, i_2, \dots, i_k)\in\NN^k$
\begin{equation}\label{eq:B-transitions}
  (q_1,\bi, q_2) \in \delta_{\tilde\A_{\NN}^k} \ \Longleftrightarrow \
  (q_1, \bb a^{i_1} \# a^{i_2}\# \dots \# a^{i_k}\eb, q_2)\in \delta_{\A}\,.
\end{equation}
The definition immediately implies that if
a word over the alphabet $\NN^k$ belongs to  $ L(\tilde\A_{\NN}^k)$, then its content belongs to $\R^k(\A)$. In the opposite direction, any relation $R\in \R^k(\A)$ is the content of a word in $L(\tilde\A_{\NN}^k)$.
Unreachable and dead states are possible in $\tilde \A_{\NN}^k$. After deleting them, we get an automaton $ \A_{\NN}^k$ having the same relations with $\R^k(\A)$. We say that a relation from  $\R^k(\A)$ is accepted by  $ \A_{\NN}^k$. Similarly, a graph $G$ is accepted by $\A$ if there exists a relation $R$ such that $G=G_R$ and $R\in L(\A_{\NN}^k) $.

To present $\A_{\NN}^k$ as a finite object, we rely on well-known facts about unary regular languages. It is clear that an automaton over the unary alphabet $\{a\}$ accepts words  $a^i$ such that $i$ is the length of an accepting run (recall that by default we assume that there are no  $\eps$-transitions). These lengths  form a 1-dimensional \emph{semilinear set}, i.e. a finite union of arithmetic progressions (a common difference of a progression is allowed to be zero).   The Chrobak-Martinez theorem  (see~\cite{Chr03, Ma02} and clarifications in~\cite{To09}) claims that, given an automaton over the unary alphabet with  $n$ states, it is possible to construct in polynomial time a~representation of the language recognized by the automaton in the form of  a~union of $O(n^2)$ arithmetic progressions $\{a +bt: t\in\NN\}$, where $a= O(n^2)$, $b=O(n)$. Unless otherwise stated, we further assume that semilinear subsets of $\NN$ are represented in this form.

For $q_1, q_2\in Q(\A_{\NN}^k)$, where $Q(\A_{\NN}^k)$ is the state set of $\A_{\NN}^k$, let $L_{q_1q_2}\subseteq \NN^k$ be the set of $x\in\NN^k$ such that $(q_1,  x , q_2)\in \delta_{\A_{\NN}^k} $.

\begin{lemma}\label{lm:Lq1q2-gen}
For $k = O(1)$, a~set  $L_{q_1q_2}$ is a union of the Cartesian products of semilinear sets, and its encoding  has the size  $\poly(\size (\A))$, where $\size(\A)$ is the size of an encoding  of $\A$.
\end{lemma}

\begin{proof}
If $\bi=(i_1, i_2, \dots, i_k)\in L_{q_1q_2}$ then there exists a run of $\A$ such that
\begin{equation}\label{eq:block-run}
q_1=q'_0\xrightarrow{\bb a^{i_1}} q'_1\xrightarrow{\# a^{i_2}} q'_2
\xrightarrow{\# a^{i_2}} \dots
\xrightarrow{\# a^{i_{k-1}}} q'_{k-1}
\xrightarrow{\# a^{i_k}\eb} q'_k =q_2
\end{equation}
(for unary relations $k=1$, thus there are no intermediate states, and~\eqref{eq:block-run} is replaced by a run $q_1\xrightarrow{\bb a^{i}\eb}q_2$, all further arguments are the same).

There are at most  $n^{k-1}$ possible tuples $(q'_1,\dots, q'_{k-1})$, where  $n$ is the number of states of  $\A$. For each tuple the set of possible values of $i_j$ is semilinear and it has a short encoding due to the Chrobak-Martinez theorem. It gives a desired representation of  $L_{q_1q_2}$.
\end{proof}

A finite encoding of  the transition relation of $\A_{\NN}^k$, i.e. the set  $\{(q_1, \bi, q_2): \bi\in L_{q_1q_2}\}$, is an encoding of the corresponding unions of Cartesian products of semilinear sets.  Lemma~\ref{lm:Lq1q2-gen} implies that the size of the representation is polynomial in the size of the encoding of  $\A$. Moreover, the encoding of the transition relation can be constructed by an encoding of $\A$ in polynomial time due to the  Chrobak-Martinez theorem.

In the main proofs we need  automata $\A$ such that  $\R^k(\A)$ is finite. We will use the following fact.

\begin{lemma}\label{lm:RkLq1q2}
  $\R^k(\A)$ is finite if and only if $L_{q_1q_2}$ is finite for all $(q_1, q_2)$. A~similar claim holds for the set of graphs accepted by  $\A$.
\end{lemma}
\begin{proof}
If $\bi\in R\in\R^k(\A)$ then $\bi$ appears on an accepting run of~$\A^k_{\NN}$. Thus it belongs to some $L_{q_1q_2}$. It implies that $R\subseteq \bigcup _{q_1, q_2}L_{q_1q_2} $, and therefore  $\R^k(\A)$ is finite if  $|L_{q_1q_2}|<\infty$  for all pairs. For  $k=2$, the finiteness of $\R^2(\A)$ implies the finiteness of the set of graphs accepted by~$\A$.

If, for some $q_1$, $q_2$, the set $L_{q_1q_2}$ is infinite, then  $\R^k(\A)$ is also infinite, since relations from  $\R^k(\A)$ contain infinitely many different $\bi\in\NN^k$. It is important here that  $\A_{\NN}^k$ has no unreachable and dead states and each transition $q_1\to q_2$ with $L_{q_1q_2}\ne\es$ belongs to an accepting run of $\A$.

Suppose that   $k=2$ and  $L_{q_1q_2}$ is infinite. Then, by Lemma~\ref{lm:Lq1q2-gen}, there exist  arithmetic progressions $A_1, A_2\subseteq \NN$ such that  $A_1\times A_2\subseteq L_{q_1q_2}$ and $|A_1\times A_2|=\infty$. Thus $(A_1\times A_2)\sm \{(i,i): i\in\NN\}$ is also infinite, since $\{i,j\}\subset A_1\cap A_2$ implies $(i,j)\in A_1\times A_2$. Therefore the set of graphs accepted by  $\A$ is also infinite.  
\end{proof}

\begin{cor}\label{cor:Rk-fin}
The finiteness of $\R^k(\A)$ is verified in polynomial time.
\end{cor}
\begin{proof}
Given $\A$, the automaton $\A^k_{\NN}$ can be constructed in polynomial time. By Lemma~\ref{lm:RkLq1q2} it is enough to check that all  $L_{q_1q_2}$ are finite. It is an easy task in the encoding format used here: one should check that in the encoding of  $L_{q_1q_2}$ from Lemma~\ref{lm:Lq1q2-gen} all the common differences of arithmetic progressions are~0.
\end{proof}

\section{A Framework for Universality Proofs}\label{sec:sketch}

A \emph{universality result} is a claim that for every nonempty language  $ X$ there exists a class  $\U_X$ of finite $k$-ary relations of the specified form such that
\begin{equation}\label{eq:univ-template}
X \leq_1 \nreg\big(\la\U_X\ra\big) \leq_2 X.
\end{equation}
Exact requirements for reductions and for the form of relations can vary. But the proofs of universality results share the same framework. We describe it in this section.

We define a function $X\mapsto \U_X$ in the following way. Denote by $ P_f(\NN^k)$ the family of all finite subsets of $\NN^k$, i.e.\ finite $k$-ary relations on $\NN$.
Choose an encoding   $\ph\colon \{0,1\}^*\to P_f(\NN^k)$ of binary words by finite relations satisfying the following requirements.
\begin{enumerate}
	\item\label{phi:req:injection} $\ph$  is an injection.
	\item\label{phi:req:pcomp} $\ph$ is  computable in polynomial time.
	\item the (partial) function $\ph^{-1}$ is also computable in polynomial time.
	\item\label{phi:req:reduction} $x\in X$ is equivalent to $\ph(x) \in \U_X$.
	\item\label{phi:req:immunity}  (The finiteness condition.) For any automaton $\A$, if $\R^k(\A)$ is infinite, then  $\A$ accepts at least one encoding $\la R\ra$ for~$R\notin \ph(\{0,1\}^*)$.
\end{enumerate}

To achieve Requirement~\ref{phi:req:reduction} we define
\begin{equation}\label{eq:def:RX}
	\U_X  = \ph(X)\cup \overline{\ph(\{0,1\}^*)}
\end{equation}
So $\U_X\cap \ph\big(\{0,1\}^*\big) = \ph(X)$ and the reduction $X \le \nreg\big(\la\U_X\ra\big)$ is defined
as $x \mapsto \A_{\la\ph(x)\ra}$, where a (deterministic) automaton $\A_{\la\ph(x)\ra}$ accepts exactly one word from the set of encodings of the relation $\ph(x)$ (reductions of this type are called \emph{monoreductions} in~\cite{Vya13}). From Requirement~\ref{phi:req:pcomp} we conclude that
\[X\leP \reg\big(\la\U_X\ra\big)\leP \nreg\big(\la\U_X\ra\big).\]

The second reduction $\nreg\big(\la\R_X\ra\big) \leq X$  is defined in different ways on automata of two types. We separate  automata, i.e. instances of  $\nreg\big(\la\U_X\ra\big)$, into  \emph{trivial} and non-trivial ones. The exact definition of triviality varies in the proofs below. In any case, $\R^k(\A) \cap \overline{\ph(\{0,1\}^*)} \neq \es$ for a trivial automaton  $\A$.
It means that  $\A$ accepts at least one word   $\la R\ra$ for a relation  $R$ that does not belong to the image of the encoding~$\ph$. An important specific case is an infiniteness of $\R^k(\A)$. In this case the finiteness condition (Requirement~\ref{phi:req:immunity}) implies $\R^k(\A) \cap \overline{\ph(\{0,1\}^*)} \neq \es$.

Due to~\eqref{eq:def:RX}, the answer in the problem  $\nreg\big(\la\U_X\ra\big)$ for a trivial automaton $\A$ is positive, in other words, 	$\A \in \nreg\big(\la\U_X\ra\big)$.
So,  the  second reduction $\nreg\big(\la\U_X\ra\big) \leq X$ on a trivial automaton sends it  to a fixed element of~$x_0\in X$. Such a definition is correct if it is possible to check triviality of an automaton using functions from a class specified for the reduction.
Formally, the disjunctive truth table reduction builds a query list consisting of the only query~$x_0$.

For non-trivial automata both answers in   $\nreg\big(\la\U_X\ra\big)$ are possible.
Let   $\A$ be a non-trivial automaton. In particular,   $|\R^k(\A)|<\infty$.
 The list of queries to the $X$ oracle consists of words $\ph^{-1}(\R^k(\A)) = \{\ph^{-1}(R): R\in \R^k(\A)\}$, where $\ph$ is the  function used in the first reduction. Therefore
\[
\A\in\nreg\big(\la\U_X\ra\big) \iff \ph^{-1}(\R^k(\A)) \cap X \neq \es,
\]
since at least one query returns  a positive answer iff $\A\in\nreg\big(\la\U_X\ra\big) $. The number of queries is finite since $\ph$ is an injection (Requirement~\ref{phi:req:injection}) and  $|\R^k(\A)|<\infty$.

Since $\R^k(\A)$ is finite, $\A^k_{\NN}$ accepts relations of the size  $\poly(\size(\A))$ only due to Lemmas~\ref{lm:Lq1q2-gen} and~\ref{lm:RkLq1q2}. All these relations can be listed using polynomial space. In some cases the reduction can be relaxed to  the   polynomial disjunctive truth table reduction   $\lettP$.

This framework should be modified for the case of invariant relations. In this case we are interested in classes of isomorphisms of finite relations, i.e. $\U_X$ should be closed under bijections on $\NN$, Thus, for a non-trivial automaton $\A$, the set  $\R^k(\A)$ is finite and all relations $\R^k(\A)$ are isomorphic to  relations from $\ph(\{0,1\}^*)$. For the second reduction we need a list of all isomorphism classes of relations in $\R^k(\A)$.
%This list can be much shorter than the total list of relations in $\R^k(\A)$.

For the universality of invariant unary relations  (Theorem~\ref{th:uuu-i} below)  the finiteness condition (Requirement~\ref{phi:req:immunity}) should be relaxed, see details in Section~\ref{sec:uuu}.

\section{Paths on Labeled Graphs}\label{sec:sets-by-graphs}

As it follows from the above plan, it is crucial to construct a list of all relations from  $\R^k(\A)$ provided the set is finite. Due to Lemma~\ref{lm:RkLq1q2}, the finiteness of $\R^k(\A)$ is equivalent to the finiteness of all $L_{q_1q_2}$, where $q_1, q_2$ are states of $\A_{\NN}^k$ (note that the number of  states of $\A_{\NN}^k$ can be less than the number of states of  $\A$, see Section~\ref{sec:automata-infinite-alphabet}).

The order of elements in a set description is insignificant. Therefore  the  characterization of relations accepted by $\A_{\NN}^k$ can be done  in terms of the following model. It  represents families of finite sets by directed graphs.

We allow loops and multiple edges in directed graphs. A \emph{transition graph} is a~tuple $G=(V,s,t,E,\ell)$, where $V$ is the set of vertices, $E$ is the set of directed edges,  $s$ is the start vertex, $t$ is the terminal vertex  and $\ell\colon E \to \{\es\}\cup U$ is the labeling function, where $U$ is a finite set called a \emph{universum of labels}.  The size $|G|$ of the transition graph is  the size of an encoding of $G$ by lists of vertices, edges and the labeling function table.

The \emph{edge label} of an edge  $e\in E$ is $\ell(e)$. For a path $P$, the \emph{label set}  $ \ell(P)$ is the union of all non-empty edge labels for all edges in the path. We also say that $P$ \emph{is marked} by $\ell(P)$. We denote by   $\F(G)$ the family of all label sets for all paths from  $s$ to $t$, $(s,t)$-paths for brevity.

Transition graphs have  a straightforward relation with automata $\A_{\NN}^k$ such that all $L_{q_1q_2}$ are finite. Vertices of~$G$  are states of $\A_{\NN}^k$ and a special terminal vertex. There is an edge from any accepting state to the terminal, this edge is labeled by the empty set. For each $\bi\in L_{q_1q_2}$ there exists an edge $(q_1, q_2)$ in the edge set of $G$, for different $\boldsymbol i$ there are different edges. This edge is labeled by~$\boldsymbol i$. The start vertex is the initial state of $\A_{\NN}^k$. It is clear that  $\R^k(\A)= \F(G)$.

The problem of verifying  $S\in \F(G)$ is algorithmically hard. At first we analyze a restricted version of the problem.

\textsc{All-labels} problem. Input: a transition graph  $G= (V,s,t,E,\ell)$. Question: does  there exist an $(s,t)$-path $P$ in $G$  such that $\ell(P)=\ell(E)$.

\begin{lemma}\label{lm:all-unordered}
  \textsc{All-labels} is  $\NP$-hard.
\end{lemma}
\begin{proof}
  We reduce SAT problem to \textsc{All-labels} problem in polynomial time. Let CNF $C = \bigwedge_{j=0}^{m-1} D_j$ be an instance of SAT.

  Define a graph and a labeling function as follows.
  The universum\footnote{We adopt notation   $[m]$ for the set $\{0,1,\dots, m-1\}$.}
  is $[m] = \{0,1,\dots, m-1\}$. A set  $S_{i,\al}\subseteq [m]$ consists of  $j$ such that  $D_j=1$ if   $x_i=\al$ (in other words, the clause $D_j$ contains the literal $x_i^{\al}$).

The vertex set of the transition graph consists of the set $X$ of variables  $x_1,\dots, x_n$ in $C$ and the start vertex $s= x_0$. The terminal is  $t=x_{n}$. Vertices $x_{i-1}$ and $x_{i}$, $0< i\leq n$, are the start and the final vertices of two paths $P_{i,0}$, $P_{i,1}$. Inner vertices of paths form disjoint sets for all paths and these sets do not include vertices from $\{x_0\}\cup X$. The length of $P_{i,\al}$ is $|S_{i,\al}|$ and $\ell(P_{i,\al}) = S_{i,\al}$ (the order of labels along the path is insignificant). There are no other vertices or edges. It is clear that this transition graph can be constructed in polynomial time.

  Note that  $(s,t)$-paths in the transition  graph constructed are in one-to-one correspondence with assignments of variables in $C$. The label set of an $(s,t)$-path corresponding to an assignment $\al$ is the set of indices of clauses that take the value 1 on this assignment. Thus, the existence of an  $(s,t)$-path marked by  $[m]$ is equivalent to the satisfiability of~$C$. It proves the correctness of the reduction.
\end{proof}

The reduction constructed in the previous proof uses only directed acyclic graphs (DAGs) without empty labels. Therefore the problem of verifying  $S\in\F(G)$ is  $\NP$-hard for DAGs as well as for the transition graphs without empty labels.

Generally, a transition graph  $G$ contains cycles and the lengths of  $(s,t)$-paths are unbounded.  Nevertheless, it is easy to see that $\text{\textsc{All-labels}}\in \NP$.

\begin{lemma}\label{lm:all-sets-quadratic}
For any transition graph $G$ and any $S\in \F(G)$, there exists an $(s,t)$-path $P$ marked by $S$ such that the  length of $P$ is $O(|V(G)|\cdot|S|)$.
\end{lemma}
\begin{proof}
  Suppose that  $S = \ell(P')$, where $P'$ is an  $(s,t)$-path. If $P' = P_1P_2P_3$ and $\ell(P_2)\subseteq \ell(P_1)$ (the subpath $P_2$ has no new labels w.r.t.\ $P_1$), then  $P_2$ can be shortened to $P'_2$ of length at most  $V(G)-1$ by deleting cycles. After this  change we have $\ell(P_1P_2'P_3) = S$.  Making the above changes one by one, we come to an $(s,t)$-path  $P$ such that $\ell(P)=\ell(P')=S$ and new labels appear along  $P$ after at most $|V(G)|$ steps.
\end{proof}

\begin{cor}\label{all-labels-NPC}
   \textsc{All-labels} is  $\NP$-complete.
\end{cor}
\begin{proof}
 Lemma~\ref{lm:all-unordered} implies that	\textsc{All-labels} is  $\NP$-hard.  Lemma~\ref{lm:all-sets-quadratic} implies that $\text{\textsc{All-labels}}\in \NP$ (a certificate is a short path marked by $\ell(E)$).  
\end{proof}

\begin{cor}\label{Gell-value}
Verifying $S\in\F(G)$ is $\NP$-complete.
\end{cor}

We present another  $\NP$-certificate for	\textsc{All-labels}. It is based on an ordered variant of the  problem. This variant is used below in the proof of Theorem~\ref{th:uuu-i}.

\textsc{All-labels-ordered} problem.
Input: a transition graph  $G= (V,s,t,E,\ell)$ such that $U=\ell(E)$ and   $U = \{u_1,\dots, u_p\}$  is linearly ordered,  $u_1<u_2<\dots<u_p$.

Question: does  there exist an $(s,t)$-path $P$ in $G$  such that $\ell(P)=U$, and the order of the  first occurrences of labels from $U$ along  the path is consistent with the order on $U$. More exactly, if  $w_1w_2\dots w_n$ is the sequence of labels along the path, the following condition should hold:
\begin{equation}\label{eq:ordered}
  \text{$w_i> w_j$ and $i<j$ imply $w_k=w_j$ for some $k<i$.}
\end{equation}
An equivalent condition: for the first occurrence of $u_j$ the prefix preceding this occurrence should be a word over the alphabet $\{u_1, \dots, u_{j-1}\}$.

\begin{lemma}\label{lm:all-ordered}
 $ \text{\textsc{All-labels-ordered}}\in\P$.
\end{lemma}

\begin{proof}
  Let  $G= (V,s,t,E,\ell)$ be an instance of \textsc{All-labels-ordered} problem.
  Construct  an automaton  $\tilde\A$ over the alphabet~$U$. The state set of  $\tilde\A$ is the vertex set   $V$ of $G$. The  initial state is $s$,
  the only accepting state is~$t$, and $(v_1, u, v_2)\in \delta_{\tilde\A}$ is equivalent to $(v_1,v_2)\in E$ and $\ell(v_1,v_2) = u$.
  The automaton $\tilde\A$ can have $\eps$-transitions. An equivalent automaton $\A$ without $\eps$-transitions can be constructed in polynomial time. Now construct  an auxiliary counting automaton $\A^c$ over the alphabet $U$.
  The state set of $\A^c$ is $Q(\A^c) = V\times[p]$, $p = |U|$. The automaton counts the number of distinct symbols appeared w.r.t.\ the order on $U$ and rejects words violating the order. The transition relation of $\A^c$ is defined as follows: $\big((q_1,i), u_j, (q_2, k)\big)\in \delta_{\A^c}$, if (a) $j\leq i=k$, $(q_1, u_j, q_2)\in \delta_{\A}$ or (b) $ j=k=i+1$, $(q_1, u_j, q_2)\in \delta_{\A}$.
The initial state is $(s,0)$, the only accepting state is $(t,p)$.

The existence of an $(s,t)$-path in $G$ marked by  $U$ is equivalent to the existence of a word  $w= w_1\dots w_n\in L(\A)$ containing all alphabet symbols. If~\eqref{eq:ordered} does hold then  $w\in L(\A^c)$. Indeed,  if  $s=q_0, q_1, \dots, q_n=t$ is an accepting run of $\A$ for $w$, then $(q_0,0), (q_1,k_1) \dots, (q_n, p)$  is an accepting run  of $\A^c$ for  $w$, here $k_i$ is the number of different symbols that are read up to $q_i$ along the run.

In the opposite direction, if $(s,0), (q_1,k_1) \dots,(q_{r-1}, k_{r-1}), (t, p)$ is an accepting run of $\A^c$ for $w= w_1\dots w_n\in L(\A^c)$, then $s, q_1, \dots, q_{r-1}, t$ is an accepting run of $\A$ for $w$, since by construction  $(q_{i-1}, w_i, q_i)\in \delta_{\A}$ for all possible~$i$. Moreover, by induction on $i$ it is easy to verify that  $k_i$ is the number of symbols of $U$ occurring in the prefix $w_1\dots w_i$ and~\eqref{eq:ordered} holds since the counter increases only when a next symbol in the order is read.
 
Thus we reduce \textsc{ALL-labels-ordered} to the non-emptiness problem for automata. The latter problem is definitely in  $\P$ because it is a variant of the reachability problem for directed graphs.
\end{proof}

We have seen that verifying $S\in \F(G)$ is  $\NP$-complete. Nevertheless, families of label sets can be efficiently generated in a sense of incremental generation with polynomial delay, in which case a next element in a set is generated in time polynomial in the input size and the number of elements found up to the moment   (see, e.g.~\cite{Gu18}). More exactly, for incremental generation of $\F(G)$ one needs to solve the following problem.

\textsc{Next-label} problem.
Input: a transition graph  $G= (V,s,t,E,\ell)$ and
a family of label sets   $\Sc = \{S_0,\dots, S_{t-1}\}$, $\Sc\subseteq \F(G)$.

Task: verify  $\F(G)  = \Sc$ and, if the answer is negative, output a set $S_t\in \F(G)\sm \Sc$.

It will be proved below that for DAGs \textsc{Next-label} problem can be solved in polynomial time w.r.t.\  the input size. Thus efficient incremental generation of  $\F(G)$ is possible if $|\F(G)|$ is polynomial in~$|G|$.

Restriction to DAGs is insignificant.

\begin{lemma}\label{th:G<DAG}
Given a transition graph $G$, it is possible to construct in polynomial time  a DAG   $G'$ such that $\F(G') = \F(G)$.
\end{lemma}
\begin{proof}
Let $G= (V,s,t,E,\ell)$, $\F(G) = \F\subseteq 2^{U}$, $|V| = n$, $|U|=m$. Given  $G$, we construct a transition graph with counter $G^c$ as follows. In $G^c$ each vertex of $G$ is covered by  $mn$ vertices and a new terminal vertex $t^c$ is added. Formally,  $V(G^c) = \big(V(G)\times [mn]\big)\cup\{t^c\}$. The edges of $G^c$ are produced from edges of $G$ by adding the increment of the counter. Formally,
\begin{multline*}
  E(G^c) = \big\{e^c: e^c =\big( (u,i), (v,i+1)\big), \ \text{where}\ 0\leq i< mn-1, (u,v)\in E(G)\big\}\cup\\
  \{ ((t,i), t^c): 0\leq i<mn\}.
\end{multline*}
The initial vertex $s^c = (s,0)$. The labeling function is defined in a natural way as
\[
\ell_{G^c}\big( (u,i), (v,i+1)\big) = \ell_G(u,v);\quad
\ell_{G^c}\big(((t,i), t^c)\big) = \es,\ 0\leq i<mn.
\]

Since the counter (the second entry of a pair determining a vertex in $G^c$) is incremented by 1 on each edge,  $G^c$ is a DAG. Paths in  $G^c$ project to paths in  $G$ by deleting the second entries of vertices. Note that a path and its projection are marked by the same label set. It implies that $\F(G^c)\subseteq \F(G)$. On the other hand, any $(s,t)$-path in  $G$ of the length at most  $mn$ can be lifted to a path in~$G^c$ by adding to the second entry of a vertex on a path in $G^c$   the number of the previous vertices along the path in $G$. It follows from Lemma~\ref{lm:all-sets-quadratic} that  $\F(G)\subseteq \F(G^c) $. Therefore $\F(G) = \F(G^c)$.

It remains to note that  $\size(G^c) = \poly(\size(G)) $ and the construction of  $G^c$ can be done in polynomial time.
\end{proof}

\begin{lemma}\label{lm:next-label}
 \textsc{Next-label} problem can be solved in polynomial time.
\end{lemma}

\begin{proof}
In fact, we use the well-known  flash-light method, see~\cite{Gu18}. Next element of a set is minimal in some suitable order. In this proof we do not need an explicit definition of the order and we skip it.

Due to Lemma~\ref{th:G<DAG}, we restrict the proof to DAGs.

In the proof we use the following notation. Let   $G= (V,s,t,E,\ell)$ be an acyclic transition graph and  $X\subseteq U$. We denote by  $G[X]$ the reduced graph, which is obtained from  $G$ by deleting all edges with labels outside  $X$ (the edges marked by empty labels remain). Edges with labels in $X$ do not change. By $w(G, S)$ we denote the number of  $(s,t)$-paths marked by a subset of  $S$. Note that  $w(G, S)$ equals the number of  $(s,t)$-paths in  $G[S]$, and this number is computed in polynomial time by the use of a~recurrence: the number of  $(s,x)$-paths is equal to the sum of the numbers of $(s,y)$-paths over  $y$ such that $(y,x)\in E(G[S])$. This equality holds since any $(s,x)$-path is split in the unique way to the beginning, $(s,y)$-path, and the last edge $(y,x)$. By  $\ld(G, S)$ we denote the number of  $(s,t)$-paths in~$G$ marked by~$S$. Thus, $\ld(G,S)>0$ is equivalent to $S\in\F(G)$. It follows immediately from the definitions that
  \begin{equation}\label{eq:w-ld}
	w(G,S) = \sum_{T\subseteq S} \ld(G,T).
  \end{equation}

Now we present an algorithm solving  \textsc{Next-label} problem. Let   $\Sc$ be a family of label sets taken from an instance of  \textsc{Next-label} problem. Add to $\Sc$ the empty set and  $U' = U\cup\{\tu\}$, where $\tu\notin U$. It results in a family~$\Sc'$. The family is partially ordered by inclusion. Construct a linear extension of the inclusion order on this family in polynomial time. (Topological sorting does solve the task.) Enumerate the sets in the family $\Sc'$ w.r.t.\ this linear extension:
\[
S_{-1} = \es,\ S_0,\ \dots,\ S_{t-1},\  S_t = U'.
\]
The algorithm computes  $\ld_{j} = \ld(G,S_j)$ in the order of increasing~$j$, and simultaneously verifies the   \emph{completeness condition for  $j$}:  $\ld(G, X) =0$ for all $X\subset S_j$, $X\notin \Sc'$. If the condition is violated, then the algorithm searches for  $X\notin \Sc'$ such that $X\subset S_j$ and $\ld(G, X)>0$. In this case, the algorithm terminates, since one more element of $\F(G)$ is found. If all completeness conditions hold, then
$\Sc'\supseteq \F(G)$. Note that  $\Sc'\sm\F(G)$ can contain $U'$ and $\es$, the latter is equivalent to $\ld_{-1}=0$.

The initial iteration corresponds to $j=-1$.  The algorithm computes in polynomial time the value of  $w(G,\es) = \ld(G, \es) =  \ld_{-1}$ in the above notation (for the empty set r.h.s. of~\eqref{eq:w-ld} contains exactly one term). If $\ld_{-1}=0$  or $\es\in \Sc$, then the completeness condition for $j=-1$ holds, and the algorithm goes to the next iteration. Otherwise, the condition is violated, which implies $\Sc\subset \F(G)$. The algorithm returns $S_t=\es\in \F(G)$.

The iteration $j$. We assume that for all $i<j$ the values $\ld_i$ are computed and $\ld_X = 0$ holds for all $X\subset S_i$, $X\notin \Sc'$, i.e. the completeness conditions hold for all $i<j$. For any set $X\subseteq U$ it is possible to compute
\[
\Delta(X) = w(G,X) - \sum_{i<j} \ld_i
\]
in polynomial time. Due to the completeness conditions for $i<j$ the equality
\[
\Delta(X) = \ld(G,X) + \sum_{Y\in \Y } \ld(G,Y)\,
\]
holds, where $\Y$ consists of proper subsets of $X$, which are not included in  $S_i$ for any $i<j$. Thus, $\Delta(X)>0$ is equivalent to the condition that there exists a subset $X$ in $\F(G)$ such that $X$ is not included in  $S_i$ for all $i<j$. Note that   $\Delta(X)>0$ and equalities $ \Delta(X\sm\{u\})=0$ for all $ u \in X$ imply that $X\in\F(G)$.

%% A.R.: Edited
The actions of the algorithm in iteration $j$ are as follows.
It constructs, a decreasing sequence of sets $S'_{j,s}$, $0\leq s \leq r$, starting from $S_j$ such that the following assertions hold:
\begin{itemize}
    \item $S'_{j,0} = S_j$ and $S'_{j,s+1} = S_{j,s}\sm \{u_{s+1}\}$,      $u_{s+1}\in S_{j,s} $;
    \item $\Delta (S'_{j,s})>0$ for each $s\leq r$;
    \item $\Delta(S'_{j,r}\sm \{u_{r+1}\})=0$ for all $u_{r+1}\in S'_{j,r} $;
    %\item $S'_{j,s+1} = S_{j,s}\sm \{u_{s+1}\}$. 
\end{itemize}
%  $S'_{j,0} = S_j$, $\Delta (S'_{j,s})>0$ for $s\leq r$, and $\Delta(S'_{j,r}\sm \{u_{r+1}\})=0$ for all $u_{r+1}\in S'_{j,r} $, and
% $S'_{j,s+1} = S_{j,s}\sm \{u_{s+1}\}$. 
%% A.R.: EoF Edited
The construction of~$S'_{j,s+1}$ implies that the next set contains one element less than the previous one.
%% Actions of the algorithm on the iteration~$j$ are the following.
%% It constructs, starting from  $S_j$, a decreasing sequence of sets  $S'_{j,s}$, $0\leq s \leq r$ such that  $S'_{j,0} = S_j$, $\Delta (S'_{j,s})>0$ for $s\leq r$, and $\Delta(S'_{j,r}\sm \{u_{r+1}\})=0$ for all $u_{r+1}\in S'_{j,r} $, and
%%  $S'_{j,s+1} = S_{j,s}\sm \{u_{s+1}\}$. The last requirement implies that the next set contains one element less than the previous one.
The construction takes at most $|S_j|^2 = \poly(|G|)$ computations of values  $\Delta(X)$: on a step~$s$ it computes $\Delta(S'_{j,s} \setminus \{u\})$ for all~$u \in S'_{j,s}$. If all these values are 0, then the process terminates; otherwise,  $u_{s+1}$ is chosen to satisfy $\Delta(S'_{j,s} \setminus \{u_{s+1}\}) > 0$.
If  $r=0$ then the completeness condition for  $j$ holds and   $\ld_j =\ld(G, S_j) = \Delta(S_j)$. Otherwise, $ S'_{j,r}\in\F(G)$ is a new element of $\F(G)$, the algorithm returns $S'_{j,r}$ and terminates.
\end{proof}

The general framework of the reduction construction from Section~\ref{sec:sketch} requires the generation of all label sets from  $\F(G)$. Due to Lemma~\ref{lm:next-label}, it can be done in polynomial time if  $|\F(G)|$ is polynomially bounded in~$|G|$.
To ensure this condition we need $\F(G)$ families with the $\eps$-separability property. We denote the symmetric difference of sets $S_1$, $S_2$ by  $S_1\oplus S_2$. A family of finite sets $\F$ has  the \emph{$\eps$-separability\/} property, if  $|S_1\oplus S_2|>\eps \cdot\max\big(|S_1|, |S_2|\big)$ for each pair $S_1\ne S_2$ of sets from $\F$ such that   $\eps\cdot \max\big(|S_1|, |S_2|\big)\geq6$.
The last requirement is needed for technical reasons, it simplifies the proofs.
It immediately implies that $\eps>0$. Note that  $\eps<2$ for any family having the $\eps$-separability property. This bound is asymptotically attained: a family of disjoint sets of the same cardinality which is at least~3,  has the $(2-\delta)$-separability property for any $2>\delta>0$.

\begin{lemma}\label{lm:distance}
  Let $n$ be the number of vertices of a transition graph $G$ and $m$ be the number of labels in its universum. If  $\F(G)$ has the  $\eps$-separability property for some $\eps$, then
  \[
  |\F(G)| \leq m\cdot n^{3/\eps}+ m^{6/\eps}.
  \]
\end{lemma}

\begin{proof}
  Since $|\F(G)|\leq 2^m$, the bound in the lemma trivially holds for $m\leq2$. So we assume in the sequel that $m\geq 3$.

  There are at most  $m^{6/\eps}$  subsets of size less than $6/\eps$ and there are at most  $m$ possible values for the size of a set from $\F(G)$.
So it is sufficient to prove that, for any  $6/\eps \leq p \leq m$, there exist at most $n^{3/\eps}$  subsets $S\in\F(G)$ such that  $|S| = p$.

Let  $d=\lfloor \eps p/2 \rfloor $, $k=\lfloor p/d \rfloor$. For an $(s,t)$-path  $P$ marked by a set $S$ with $|S|=p$, define a sequence of vertices	$\bv(P)=(v_0, v_1, \dots, v_{k}, v_{k+1})$, which is a subsequence of vertices of  $P$, as follows. Set $s=v_0$, $t= v_{k+1}$. Let $P_i$ be the initial segment of  $P$  finishing at the vertex $v_i$. We require that  $|\ell(P_i)\sm \ell(P_{i-1})| = d $ for all  $0<i\leq k$. The choice of vertices  $v_i$ is possible, since each edge adds to the label set at most one label. Due to the choice of parameters, we also have $|\ell(P)\sm\ell(P_k)|<d$.

Now we prove that  $\ell(P)\ne \ell(P')$ implies $\bv(P)\ne \bv(P')$. The lemma immediately follows from this fact, since the number of distinct sequences  $\bv(P)$ is at most $n^k$ and $k\leq p/d \leq 3/\eps$ provided $\eps p\geq6$.

Our proof is contrapositive:  $\bv(P)= \bv(P')$ implies $\ell(P)= \ell(P')$.
Let $S = \ell(P)$, $S' = \ell(P')$, $S_i =\ell(P_i)\sm \ell(P_{i-1}) $, $S'_i = \ell(P'_i)\sm \ell(P'_{i-1})$. Again,  $P_i$ ($P'_i)$ is the initial segment of $P$ ($P'$) finishing at  $v_i$. By induction on $i$ we prove that $S_i\subseteq S'$, the inclusions $S'_i\subseteq S$ are proved similarly.

The induction base $i=1$.  Change the first segment of $P'$ by the first segment of  $P$. It gives a new $(s,t)$-path $P''$ marked by a set $S''$ such that  $ S_1\cup (S'\sm S'_1)\subseteq S'' \subseteq S_1\cup S'$. These inclusions imply   $S''\oplus S'\subseteq S_1\cup S'_1$. We conclude that  $|S''\oplus S'|\leq 2d\leq \eps p$. Due to the $\eps$-separability property, we get $S'' = S'$, thus $S_1\subseteq S'$.

The induction step. Suppose that  $S_i\subseteq S'$ for all $i< j$. Change the $j$-th segment of   $P'$  by the $j$-th segment of~$P$. It gives a new  $(s,t)$-path $P''$ marked by a set $S''$.
 The inclusions  $ S_j\cup (S'\sm S'_j)\subseteq S'' \subseteq S_j\cup S'$ also hold in this case.  Note that on the $j$-th segment of $P$ not only labels from $S_j$ can appear, the labels from  $S_i$, $i<j$, are also possible.   But all such labels are in $S'$ by the inductive hypothesis.
Repeating the arguments from the proof of the induction base, we conclude   $S'' = S'$, which implies $S_j\subseteq S'$.
\end{proof}

We also need an algorithmic variant of Lemma~\ref{lm:distance}.

\begin{lemma}\label{lm:distance-alg}
  Let  $\C$ be a family of finite sets such that   $\C\in\P$ and for any transition graph $G$ it follows from $\F(G)\subseteq \C$ that $\F(G) $ has the $\eps$-separability property for some  $\eps = \eps(G)$, where $\eps(G)$ is computable in polynomial time.
Then there exists an algorithm such that it takes  a transition graph  $G$ as input,  verifies the inclusion   $\F(G)\subseteq \C$, and, if the inclusion holds,  generates a list of all sets of	$\F(G)$. The running time of the algorithm on input $G$ is $\poly(|G|^{O(1/\eps(G))})$.
\end{lemma}
\begin{proof}
  Let  $n$ be the number of vertices of a transition graph $G$, and $m$ be the number of labels in a universum $U$. Assume that $\F(G)$ has the $\eps$-separability property.

  The algorithm applies the procedure of solving \textsc{Next-label} problem from Lemma~\ref{lm:next-label}, starting from the empty family $\Sc$. For each new set $S_t$ generated by the procedure, the algorithm verifies  $S_t\in \C$. In the case of a negative answer the algorithm immediately terminates with the answer $\F(G)\nsubseteq \C$. If more than  $m\cdot n^{3/\eps}+ m^{6/\eps}$ sets is generated, then the algorithm also terminates with the same answer. The answer is correct due to Lemma~\ref{lm:distance}. Note that if  $\F(G) \nsubseteq \C$, then at some point the generation algorithm will generate a set $S_t \not\in \C$, but it can happen  too late. This rule makes the running time shorter.

If the whole set	$\F(G)$ is generated, then  $\F(G)\subseteq \C$, since for all sets in  $\F(G)$ it is verified that they are in~$\C$.

The running time of the algorithm is
\[(m\cdot n^{3/\eps}+ m^{6/\eps})\cdot\poly(m,n) = \poly(|G|^{O(1/\eps)}).\qedhere\]
\end{proof}

To use this construction in universality proofs, we need \emph{large} families having the $\eps$-separability property, i.e.\ the families of size exponential in the size $m$ of a~universum. Moreover, Lemma~\ref{lm:distance-alg} requires that these families should be decidable in polynomial time. To satisfy these requirements,  we use efficient asymptotically good codes.

Recall the basic definitions in  coding theory. A linear binary $(n, k, d)$-code $C$ is a subspace of the coordinate vector space $\FF_2^n$ of dimension  $k$ such that each nonzero vector in $C$ has at least  $d$ entries are~$1$.  A code is called asymptotically good if $k = \Omega(n)$ and $d = \Omega(n)$. The asymptotic notation implies that we are defining a~family of codes with $n\to\infty$.   For our purposes it is convenient to parametrize families of codes by the code dimension $k$ instead of the block length. So we denote codes from a family as $C_k$. We need efficiency in a weak sense. Namely, there exists a polynomial time encoding algorithm that computes  linear maps $c_k\colon \FF_2^k\to\FF_2^n$ such that $c_k(\FF_2^k) = C_k$ and the function $f(k,x) = c_k(x)$ is computable in time $\poly(k,n)$. More details about coding theory can be found in~\cite{McWSl, RRShen}.

It is well-known that efficient asymptotically good codes do exist, see, e.g.,~\cite{McWSl,Sp96, CRVW02, Gu04, RRShen}.  For our purposes it is convenient to parametrize families of codes by code dimension $k$ instead of the block length. From now on, we fix an efficient family of codes $C_k$ with parameters $(r k, k, \geq k)$, where $r$ is an absolute constant. We assume that  $r\geq4$.  W.l.o.g.\ also assume that  $C_k$ is not contained in a smaller coordinate space. In other words, no coordinate is identically zero on  $C_k$.

Using a code   $C_k$ we define a $1/(r+1)$-separable family  $\C_M$ of nonempty subsets of $[M]$, $M= r k$. Let  $0\ne c_k( x)\in \FF_2^M$ be a nonzero code vector.   The corresponding subset~$I_x\subseteq [M]$ consists of integers $i$ such that $c_k( x)_i=1$. The addition of  $M$-dimensional  vectors corresponds to the symmetric difference of the corresponding sets. Thus, from the code distance bound, we get the  $1/(r+1)$-separability property for  $\C_M$ provided  $k\geq 6(r+1)$, i.e.\ $M\geq 6r(r+1)$. Extra 1 is needed because in the definition of the separability property we use the strict inequality.
The family $\C_M$ consists of  $2^{M/r}-1$ subsets each of size at least $M/r$, since the code distance is at least $M/r$.

Note that the universum can be changed by any set of the same size in the problem of representing subsets by transition graphs.
Below we also denote by $\C_M$ the families that are isomorphic to $\C_M$ to simplify notation. The exact meaning of $\C_M$ will be clear from the context. The following two constructions of this kind will be used later.

Sets in an infinite family  $\C_\infty$ consist of binary words. This family is a disjoint union of families isomorphic  $\C_M$ for all $M$. Let  $M = r k$ and $s = \lceil \log M\rceil$. (All logarithms in the text are binary.)
Define an injection $[M]\to \{0,1\}^s $ by the rule:  the image of $0\leq i< M$ is a binary representation of $i$ of length  $s$ (appended by zeros to the left, if needed). Since  $r\geq4$, different values of $M$ have different corresponding values of $s$ and the families $\C_M$ do not intersect. Note that  $\C_\infty\in \P$, therefore Lemma~\ref{lm:distance-alg} can be applied to it.  

If we need to embed binary words (i.e. elements of sets in $\C_\infty$)  into nonnegative integers, then we use the standard bijection. A word  $w\in \{0,1\}^*$ is mapped to an integer $\nu(w)$, which is obtained from an integer represented in binary as  $1w$ by subtracting~1. It is clear that this bijection is computable in polynomial time.

\section{Universality of Unary Relation Descriptions}\label{sec:ubu}

\begin{theorem}\label{th:ubu}
For each language $\es\subset X \subseteq \{0,1\}^*$, there exists a family $\U_X$ of finite subsets of $\NN$ such that
\[
X \leP \nreg\big(\la\U_X\ra\big) \lettNP X.
\]
\end{theorem}

\begin{proof}
We follow the plan outlined in Section~\ref{sec:sketch}. The first step is to define an appropriate encoding  $\ph\colon \{0,1\}^*\to P_f( \NN)$.
Take a binary word $w$ of the length $k$ and construct the corresponding code word $w' = c_k(w)\in C_k$ of the length $M= rk$. It takes a time polynomial in $k$.  Next, construct the corresponding subset of words $S_{w'}$ from the family $\C_M$ as it described in Section~\ref{sec:sets-by-graphs}. There are at least $k$ words in  $S_{w'}$ (the code distance of $C_k$ is at least $k$).  Their lengths are  $s = \lceil \log rk\rceil$.  Thus  $|S_{w'}|= \poly(k)$.  Finally,   $\ph(w)\in P_f(\NN)$ is defined as  $\nu(S_{w'})$, where the standard bijection between non-negative integers and binary words  $\nu$ is applied to each element of $S_{w'}$.  All integers in  $\ph(w)$ belong to the range from $2^{s}-1$ to $2^{s+1}-2$.  Thus the maximum of them is at most 2 times larger than the minimum of them.  This property of $\ph$ is important for the finiteness condition.  The image $\ph( \{0,1\}^*)$ is decidable in polynomial time, since $\nu^{-1}$ is computable in polynomial time as well as the check $\nu^{-1}(S)\in \C_M$ (note that  $\nu^{-1}(S)$ determines the value $M=rk$, since $r\geq4$).
On the image $\ph(\{0,1\}^*)$, the inverse map is correctly defined and it is computed in polynomial time, since computing $w'$ of the length $M$, corresponding to $\nu^{-1}(S)$, the check $w'\in C_k$, and computing $w= c_k^{-1}(w')$ can be done in polynomial time.

The family  $\U_X$ is defined as in Section~\ref{sec:sketch}:  $\ph(X)\cup \overline{\ph(\{0,1\}^*)}$.

Since $\ph$ is polynomially computable, we immediately get a monoreduction that maps  $w$ to the NFA $\A_{w}$ that recognizes a single word from the set of descriptions of $\ph(w)$. So, the first part of the theorem is proved.

Now we prove the finiteness condition. Suppose that  $\R^1(\A)$  is infinite for an automaton $\A$. Due to Lemma~\ref{lm:RkLq1q2} at least one of the languages  $L_{q_1q_2}$ defined for the automaton $\A^1_\NN$ is infinite too. As usual, we assume that there are no unreachable and dead states. It implies that $\R^1(\A)$ contains sets in the form  $S\cup\{t\}$, , where $S$ is fixed and $t$ is arbitrarily large. Some of these sets do not belong to $\ph(\{0,1\}^*)$, hence they belong to $\U_X$. Indeed, if $S\ne\es$ then, in   $S\cup\{t\}$ for sufficiently large $t$, the maximum of elements of the set at least 3 times larger than the minimum, such a set does not belong to $\ph(\{0,1\}^*)$.  If $S=\es$ then, for sufficiently large $t$ the set $\{t\}$ does not belong to $\ph(\{0,1\}^*)$, since there are at least 2 elements in a set from $\ph(\{0,1\}^*)$ that contains $t$ for sufficiently large $t$.

A trivial automaton in this case accepts at least one encoding of a set from  $\overline{\ph(\{0,1\}^*)}$.
In particular, if $|\R^1(\A)|=\infty$ then $\A$ is trivial. Due to Corollary~\ref{cor:Rk-fin} it takes a polynomial time to check that $|\R^1(\A)|=\infty$. If $|\R^1(\A)|<\infty$, then a set from $\R^1(\A)$  has size $\poly(\size(\A))$ due to Lemma~\ref{lm:Lq1q2-gen}.
Therefore the list of elements of $\R^1(\A)$ can be generated in polynomial time due to Lemma~\ref{lm:next-label}.
Since $\ph(\{0,1\}^*)$ is decidable in polynomial time, the condition $\R^1(\A)\cap \overline{\ph(\{0,1\}^*)}\ne\es$  can be verified in polynomial time.

The second reduction   $\nreg\big(\la\U_X\ra\big) \lettP X$ sends trivial automata to a fixed element of~$X$. For a non-trivial automaton it constructs a family of sets from  $\R^1(\A)$ and generates the corresponding list of queries to the $X$ oracle, as it explained in Section~\ref{sec:sketch}. In this way we get a disjunctive truth table reduction in polynomial time.
\end{proof}

\section{Universality of Invariant Binary Relation Descriptions}\label{sec:bbu}

In the analysis of invariant graph properties  the edge labels are pairs of non-negative integers and the label sets are binary relations. A transition graph $G$ defines a family of graphs   $\Fg(G)$ called \emph{label graphs}. By definition, a label graph $F$ belongs to $\Fg(G)$ if  $F = G_R $ for  $R\in \F(G)$ (the definition of $G_R$ see in Section~\ref{ssec:rel-descriptions}, Equation~\eqref{eq:def:GR}). The label graphs are considered up to isomorphism. For brevity, we say in this section `a label graph $H$' instead of `an isomorphism class containing $H$'.

We encode binary words by graphs in the following way. A~binary word $w= w_1w_2\dots w_{n}$ corresponds to the graph $\ph(w) = H_w$, which is produced from a path  $v_0v_1\dots v_{n+1}$ by attaching a clique of size 6 to the vertex~$v_0$,  attaching a clique of size 5 to the vertex $v_{n+1}$, and attaching a pendant edge $v_iv'_i$ to  $v_i$, where $1\leq i\leq n$, provided $w_i=1$. It is clear that  $\ph (w)$ is computable in polynomial time. Moreover, it is an injection to isomorphism classes due to the following proposition.

\begin{prop}\label{GS-nonisomorhic}
  If $H_w \cong H_{w'}$ then $w=w'$.
\end{prop}
\begin{proof}
  Suppose that   $H_w \cong H_{w'}$. Both graphs contain disjoint cliques of sizes 6 and 5, and these cliques contain vertices of degrees $6$ and $5$ respectively. The graph induced by that pair of vertices and vertices outside the cliques is a caterpillar, i.e.\ a~tree  such that  deletion of pendant vertices transforms it to a path. Degrees of non-pendant vertices of this caterpillar are 2 or 3 by construction. An isomorphism between $H_w$ and $H_{w'}$ sends the cliques $K_6$ and $K_5$ to the corresponding cliques and the caterpillar is sent to the corresponding caterpillar.

  Since degrees of  vertices are preserved by an isomorphism, we get  $w=w'$.
\end{proof}

Denote by  $\G$ the image $\ph(\{0,1\}^*)$ of the whole set of binary words. Graphs in  $\G$ are efficiently recognized.

\begin{lemma}\label{lm:Giso}
Given a graph $G$, the isomorphism of $G$ and a graph from $\G$  can be verified in polynomial time. In the case of positive answer, a word $w\in \{0,1\}^*$ such that  $G\cong H_w$ can be found in polynomial time.
\end{lemma}
\begin{proof}
  Verify the properties used in the previous proof of Proposition~\ref{GS-nonisomorhic}.  If any property is violated, return the negative answer.

  Check that vertices of degree greater than 3 induce a pair of disjoint cliques $K_6$ and $K_5$. Find in the cliques vertices of degrees 6 and 5 respectively. Check that the graph induced by these vertices and vertices of degree less than 4 is a caterpillar. Degrees of non-pendant vertices of the caterpillar are 2 or 3. Restore $w$ by these degrees.
\end{proof}

\begin{theorem}\label{th:bbu}
For every language $\es\subset X\subseteq \{0,1\}^*$  there exists a set $\G_X$ of graphs closed under isomorphisms such that
\[
X \leP \nreg\big(\la\G_X\ra\big) \leGen{dtt}{PSPACE} X.
\]
\end{theorem}

\begin{proof}  
Again, we follow  the general plan outlined in Section~\ref{sec:sketch}.  To encode binary words by graphs, we use the above encoding $\ph(w)$.
 Lemma~\ref{lm:Giso} implies that, by any graph isomorphic to $\ph(w)$, a word $w$ is restored in polynomial time.
 The class $\G_X$ is a union  of graphs isomorphic to graphs $\ph(w)$, $w\in X$ and graphs that are non-isomorphic to any $\ph(w)$, $w\in\{0,1\}^*$.

Now we prove the finiteness condition for  $\G_X$. Suppose that $\R^2(\A)$ is infinite for an automaton $\A$. Lemma~\ref{lm:RkLq1q2} implies that at least one language $L_{q_1q_2}$ is infinite for the automaton $\A^2_{\NN}$. By construction, there are no unreachable and dead states in  $\A_{\NN}^2$. Thus, for any accepting path of  $\A^2_{\NN}$ containing the transition from  $q_1$ to $q_2$, there exists a pair $(i,j)\in L_{q_1q_2}$ such that  $i$ (or $j$) is not read on all other transitions. It means that there exists a vertex of degree 1 in the label graph marking this path. This graph belongs to $\G_X$, since graphs from  $\G$ have no vertices of degree~1.

A trivial automaton $\A$ in this case is an automaton such that $|\R^2(\A)|= \infty$ or $\A^2_{\NN}$ accepts a graph that is not isomorphic to graphs from the family~$\G$.
The first condition is verified in polynomial time due to Corollary~\ref{cor:Rk-fin}.
The second one can be verified in polynomial time by an algorithm with an $\NP$-oracle. A~certificate is a~graph accepted by $\A^2_{\NN}$ but non-isomorphic to any graph in $\G$.  Since the size of a graph accepted $\A^2_{\NN}$ is polynomially upperbounded by the size of $\A$, and it takes a polynomial time to check the isomorphism to a graph from  $\G$ by Lemma~\ref{lm:Giso}, the correctness of a certificate is verified in polynomial time. Thus, the second condition is verified in polynomial space.

For trivial automata the answer in $\nreg\big(\la\G_X\ra\big) $ is positive. The second reduction sends all trivial automata in some fixed element of~$X$.

If $\A$ is non-trivial, then  $\R^2(\A)$ is finite and $\A^2_{\NN}$ accepts only graphs isomorphic to graphs from~$\G$. The second reduction generates a list of isomorphism classes of graphs accepted by   $\A^2_{\NN}$. It is possible in polynomial space, since all label graphs have polynomial size in the size of $\A$.
Then the reduction computes $\ph^{-1}$ for each graph in the list and put the result into a list  of queries to the $X$ oracle. It gives the required reduction  $\nreg\big(\la\G_X\ra\big)\leGen{dtt}{PSPACE} X$.
\end{proof}

\section{Universality of Invariant Unary Relation\\ Encodings}\label{sec:uuu}

The size is the  only invariant of a finite set of $\NN$ under bijections of $\NN$. Therefore an invariant family $\I_L\subset 2^{\NN}$ of unary finite relations on $\NN$ is completely determined by a subset $L\subseteq \NN$ and consists of finite subsets of $\NN$ such that the size of a~subset belongs to~$L$. It limits the ability to encode binary words since, for words of the length  $n$, we need exponentially large integers. Recall that we represent integers in unary. So in this case we prove a weaker universality  result for  the unary alphabet. This result is interesting because it shows how to prove universality results when the finiteness condition does not hold.

\begin{theorem}\label{th:uuu-i}
For any  $\es\subset X\subseteq\{a\}^*$ there exists a set $L\subset \NN$ such that
\[
X \leP \nreg\big(\la\I_L\ra\big) \leTNP X.
\]
\end{theorem}

\begin{proof}
Languages over the unary alphabet are in one-to-one correspondence with subsets of $\NN$ (a word corresponds to the length of the word).
 So we assume that  $X\subset \NN$ and, for algorithmic statements, assume that integers are represented in unary. Let  $\ph(x) = \binom{x+1}2$ and  $L= \ph(X)\cup\overline{\ph(\NN)}$.

It is clear that $\ph$ is computable in polynomial time as well as the inverse of $\ph$, and $\ph(\NN)$ is decidable in polynomial time.

The first reduction in the theorem is a~monoreduction. A~word
\[
w = \bb a\eb \bb a^2\eb \dots \bb a^{\ph(x)}\eb
\]
is an encoding of a set of the size~$\ph(x)$. We reduce  $x$ to an automaton $\A_{w}$ recognizing the language $\{w\}$.

It seems that it is impossible to satisfy the finiteness condition in this case.
But a weaker condition holds.  Namely, if there exists a cycle in the graph of an automaton  $\A^1_\NN$ and  an edge $(q_1, q_2)$ on the cycle such that  $L_{q_1q_2}$ is infinite (the corresponding transition is called an $\infty$-transition), then  $\R^1(\A)$ contains a set from $\I_L$. We call such automata trivial in this case. This weaker condition can be verified in polynomial time, since it is reduced to the reachability problem in a~directed graph.

To prove this weaker condition, suppose for the sake of contradiction that	$L_{q_1q_2}$ contains an infinite arithmetic progression $a + b x$, $b>0$, $x\geq 0$. Recall that by construction $\A^1_\NN$ has no unreachable and dead states.   Choose an accepting path that goes  $t$ times along the cycle containing a transition from  $q_1$ to $q_2$.  One can choose at each time a~sufficiently large integer from the progression and the same integers on the rest of transitions. It results in  accepting paths for sets of the sizes $N+t$ for a suitable constant $N$.  It is possible to choose  $t$ such that  $N+t\notin\ph(\NN)$, therefore it belongs to~$\I_L$.

The second reduction sends all trivial automata to a fixed element of~$X$, since for these instances the answer is positive in the problem $\nreg\big(\la\I_L\ra\big)$.

Now consider non-trivial automata.  Let $n$ be the number of states of an automaton $\A$, which is the input for the reduction, and $m$ be the number of transitions $q_1\to q_2$ of $\A^1_{\NN}$ such that  $L_{q_1q_2}$ is finite.  Each accepting run of $\A^1_{\NN}$ contains at most  $n$ $\infty$-transitions, otherwise some of them belong to a cycle.
Other transitions $q_1\xrightarrow{} q_2$ can belong to a cycle. Thus a set accepted on this run can contain many elements from   $L_{q_1q_2}$. But the set  $L_{q_1q_2}$ for these transitions is finite and its size is upperbounded by  $O(n^2)$ due to Chrobak-Martinez theorem. Therefore $\A^1_\NN$ accepts sets of size $O(mn^2)$.

The list   $\ell_1, \dots, \ell_s$ of the sizes of sets accepted by  $\A^1_\NN$ can be generated in polynomial time by an algorithm with an  $\NP$-oracle. Note that $\max_i \ell_i = O(mn^2)  =O(\poly(\size(\A)) $. To generate a list, an algorithm checks, for each possible value of~$\ell$, that $\A^1_\NN$ accepts a set of the size  $\ell$ (if so,  $\ell$ is included in the list).

 Suppose that an accepting run   $q_0q_1\dots q_p$ of  $\A^1_\NN$ makes transitions reading integers
\[
x_1, x_2, \dots ,x_p
\]
and there are exactly $\ell$ distinct among them.  List them in the order of occurrence on the accepting path. It gives a sequence  $r = (r_1,r_2,\dots, r_\ell)$.

Define an automaton $\B(r)$ over the alphabet  $\{r_1,\dots, r_\ell\}$  having the same states, the same initial state, and the same accepting states as $\A_{\NN}^1$ has.
The transition relation define as follows
\[
\begin{aligned}
  &(q_1,r_i, q_2)\in \delta_{\B}&\text{is equivalent to}\ &(q_1,   r_i, q_2)\in \delta_{\A^1_\NN}.
\end{aligned}
\]

It is clear from the construction that  $\A^1_\NN$ accepts a set of size~$\ell$ if and only if  there exists a word in $L(\B(r))$ such that all the symbols $r_1$, $r_2$, $\dots$, $r_\ell$ occur in the word and they appear in the order of their indices.

Lemma~\ref{lm:all-ordered} can be used to check efficiently the existence of a word containing symbols $r_i$ in the order fixed in the sequence $r$.

But there is another difficulty in construction of $\B(r)$. Some $r_i$ can be very large if they appear on  $\infty$-transitions. Therefore, we represent integers $r_i>n^3$ by indicating an $\infty$-transition  $q_1\xrightarrow{} q_2$ of $\A^1_\NN$  and an infinite  arithmetic progression  $a+ bt$, $a=O(n^2)$, $ b = O(n)$ in $L_{q_1q_2}$. We assume that the progression is taken from succinct description of a regular language over the unary alphabet from the Chrobak-Martinez theorem, as it is done in Lemma~\ref{lm:Lq1q2-gen}.

Using the certificate construction algorithm, it is possible to construct  $\B(r)$ in polynomial time by an algorithm with an $\NP$-oracle, if the correctness problem for descriptions of it belongs to $\NP$. An instance of the correctness problem is an automaton $\B$ with the state set  $Q(\A_{\NN}^1)$ over the alphabet $\{r_1,r_2,\dots, r_\ell\}$, each transition of $\B$ corresponds to a transition of  $\A^1_\NN$ and is marked by an integer or an infinite arithmetic progression as explained above.
In addition to the occurrence order, the integers $r_i$ are ordered by their values (note that the orders can differ).  The correctness of labels can be verified in polynomial time due to the Chrobak-Martinez theorem.
The existence of a word in  $L(\B)$ such that all alphabet symbols occur in the word in the prescribed order can be verified in polynomial time using Lemma~\ref{lm:all-ordered}.  There exists one more correctness condition:  $r_i$ can appear on several $\infty$-transitions, thus it should be possible to substitute each $r_i$ by an integer in such a way that for different $r_i$ integers are different.  Since the order of values for $r_i$ is given, it is a~particular case of ILP (a satisfiability problem for a system of linear inequalities in integer-valued variables). It is well-known that ILP is in  $\NP$~\cite{Schrijver}. Therefore this last correctness condition can be verified by an  $\NP$-oracle.

To complete the proof we note that if
\[
\{\ell_1,\dots,\ell_s\}\cap \overline{\ph(\NN)}\ne \es,
\]
then  $\R^1(\A)\cap \I_L\ne \es$.  In this case  reduce $\A$ to a fixed element of $X$. Otherwise, generate the list $\big(\ph^{-1}(\ell_1), \dots, \ph^{-1}(\ell_s)\big)$. In that case $L(\A^1_\NN)\cap \I_L\ne \es$ is equivalent to $\ph^{-1}(\ell_i)\in X$ for at least one~$i$. The last condition can be verified by a call of $X$ oracle on the list.
\end{proof}

\section{Concluding Remarks}

In this paper we extend a list of filter classes having the universality property. The previously known examples are arbitrary and prefix-closed filters~\cite{Vya13}, and filters that are languages of correct protocols of work of finite automata  equipped with an auxiliary data structure~\cite{RV22}. To compare new filters to the existing ones, we provide several remarks. At first, stronger restrictions on filters results in stronger reductions, which give weaker universality properties. It is quite natural. Second, proofs of universality properties for new filter classes use more sophisticated tools. In particular, efficient asymptotically good  codes play  an important role in the proofs. It is a peculiar situation in Formal Language Theory. We know one more example of results of this kind. In~\cite{ChV20} quasi-polynomial lower bounds on the translation from one-counter automata to Parikh-equivalent nondeterministic finite automata were proved with use of error-correcting codes. Note that for proofs in~\cite{ChV20} it is sufficient to use codes that are not asymptotically good. Moreover, to prove universality property we develop a new model of representing sets by paths in directed graphs. This new model of non-uniform computation differs in many properties from the standard ones. We consider this model as interesting on its own and hope that it will be useful in other applications too. At third,  all known examples of filter classes possessing the universality property, despite severe structural limitations, are parametrized by arbitrary languages or similar objects (say, families of finite relations are used in this paper).

Is it possible to put so strong structural limitations on filters parametrized by arbitrary languages that the corresponding filter class does not have the universality property? We mean here the weakest variant of the property using general Turing reductions. The answer is unknown to us. It seems plausible that an example of this sort can be constructed using  tools from algorithm theory. But such a construction would look artificial.

The second interesting question is about the complexity of reductions. Does the universality property for invariant classes of graphs hold with respect to polynomial disjunctive truth table reductions? Polynomial Turing reductions?
It would be quite interesting to answer this question even assuming the standard complexity-theoretic conjectures. In our opinion,  results of this sort require new technical tools.

\paragraph*{Acknowledgments.}
The authors are thankful to anonymous referees for valuable remarks and suggestions on improvement of this paper.
%% This work is supported by the Russian Science Foundation grant 20--11--20203.

\end{document}